# Stability and electronic properties of "4-8"-type ZnSnN$_2$ thin films free of spontaneous polarization for optoelectronic devices

D. Q. Fang[1*]

[1]MOE Key Laboratory for Nonequilibrium Synthesis and Modulation of Condensed Matter, School of Physics, Xi'an Jiaotong University, Xi'an 710049, China

## Abstract

Ternary nitride ZnSnN$_2$ is a promising photovoltaic absorber material. In this work, using first-principles calculations, we investigate the stability and electronic properties of "4-8"-type ZnSnN$_2$ thin films. We find that below a certain thickness "4-8"-type thin films have lower total energy than polar films. For 4-layer ZnSnN$_2$ thin film, the Pna2$_1$/Pmc2$_1$ → 4-8 transition can spontaneously occur at finite temperatures. All "4-8"-type thin films studied are semiconducting and free of spontaneous polarization, the bandgaps of which can be tuned by the thickness of films, ranging from 1.4 eV to 1.8 eV. Furthermore, these films show light electron effective masses, and octet-rule-preserving disorder has insignificant effects on the electronic properties. Our results provide new insights into the structure of ZnSnN$_2$ in the thin film form and guidance for the experimental investigation.

[*]fangdqphy@xjtu.edu.cn

# I. Introduction

Ternary Zn-IV-N$_2$ nitrides with IV = Si, Ge, and Sn have attracted considerable attention due to the tunability of functionalities and earth abundance of constituent elements. They can be thought of as analogs of wurtzite III-N nitrides, with the group III element replaced alternately by a group II and a group IV. By the substitution, the structure changes from wurtzite to orthorhombic (space group Pna2$_1$) structures. [1,2] ZnSiN$_2$ and ZnGeN$_2$ have been investigated as wide-bandgap semiconductors with predicted bandgaps of 5.4 eV and 3.4 eV, respectively. [3] For high-quality single crystalline ZnSnN$_2$ films, the optical bandgap was determined to be $1.67 \pm 0.03$ eV. [4] Zn-IV-N$_2$ nitrides and their alloys possess a wide range of attainable bandgaps [5–7], similar to the III-N, having potential applications in optoelectronic devices.

In particular, ZnSnN$_2$ is regarded as one of the promising absorber materials for photovoltaic applications due to its suitable bandgap, robust optical absorption, and low cost [8]. Its experimental synthesis was achieved using different techniques such as radio-frequency sputtering [1], molecular beam epitaxy [9], and plasma-assisted vapor-liquid-solid growth [10]. We refer the reader to Refs. [4,11–13] for the research progress of ZnSnN$_2$ growth.

According to the theoretical calculations of Ref. [14], the energetically most favorable structure of ZnSnN$_2$ is the Pna2$_1$ orthorhombic structure, while the Pmc2$_1$ orthorhombic structure is higher in total energy by $13 \pm 3$ meV per formula unit than the Pna2$_1$. X-ray diffraction measurements showed that ZnSnN$_2$ epitaxial films on ZnO substrates with growth along the [0001] direction have a wurtzite crystal structure [4,15]. Nevertheless, the bandgap of ZnSnN$_2$ determined from the optical measurements is in close agreement with the predicted ones based on perfectly ordered Pna2$_1$ crystals [1,14,16]. To reconcile these seemingly contradictory observations, Quayle *et al*. [14] proposed an octet-rule-preserving model of disorder that can be viewed as combinations of the row stackings in the basal plane occurring in the Pna2$_1$ and Pmc2$_1$ structures, which yields the measured, wurtzite-like X-ray diffraction spectra and predicts a bandgap that is relatively insensitive to ordering.

Although the bulk properties of ZnSnN$_2$ such as elasticity, cation disorder, bandgap, and defects have been extensively investigated by first-principles calculations [14,16–18], theoretical studies on ZnSnN$_2$ thin films are comparatively limited. Karim *et al*. designed an InGaN-ZnSnN$_2$ based quantum well structure with a thin (nanometer-scale) ZnSnN$_2$ layer inserted into InGaN quantum well, which can be used to realize high-efficiency amber ($\lambda \sim$ 600 nm) light emitting diodes [19]. Our group proposed monolayer ZnSnN$_2$ in a graphene-like planar structure as visible-light photocatalyst [20]. Subsequently, similar monolayer structures for Zn-IV-N$_2$ (IV = Si, Ge, and Sn) [21] and Cd-IV-N$_2$ (IV = Si and Ge) [22] compounds were predicted to be stable. On the other hand, for wurtzite binary materials like GaN and ZnO, below a certain thickness, the wurtzite polar films with exposed {0001} surfaces are not stable, and the body-centered-tetragonal (BCT) structure is found to be favored polymorph [23–25]. Considering the similarity between the wurtzite and Pna2$_1$ orthorhombic structures, it is interesting to investigate the structural stability of ZnSnN$_2$ thin films and explore their electronic properties for possible device applications.

## II. Computational details

Our first-principles calculations are performed using density functional theory as implemented in the Vienna Ab initio Simulation Package (VASP) code. [26] The exchange-correlation potential is described within the generalized gradient approximation of Perdew, Burke, and Ernzerhof (PBE). [27] The electron-ion interaction is described by the projector augmented-wave method. [28,29] The plane wave energy cutoff is set to 500 eV. A 4×4×4 (8×4×4) Monkhorst-Pack **k**-point grid [30] is employed for the primitive cell of Pna2$_1$ (Pmc2$_1$) bulk structure and commensurate **k**-point grid is used for thin films. A vacuum region of more than 11 Å is employed to avoid the interactions between neighboring slabs. For polar slabs, dipole correction is exerted in the out-of-plane direction. [31] The electronic self-consistency and ionic forces are converged to $10^{-5}$ eV and $10^{-2}$ eV/Å, respectively. To predict reasonable bandgap, we employ the modified HSE06 hybrid functional [32] with the fraction of the nonlocal Fock exchange set at 0.30 (denoted as *mod*-HSE06).

## III. Results and Discussion

### A. Bulk properties of ZnSnN$_2$

We first investigate the bulk properties of ZnSnN$_2$ in the Pna2$_1$ and Pmc2$_1$ structures (designated as Pna2$_1$- and Pmc2$_1$-ZnSnN$_2$) with 16 and 8 atoms in the primitive cells, respectively (see APPENDIX). The optimized lattice parameters are predicted as $a$ = 6.811 Å, $b$ = 5.909 Å, and $c$ = 5.540 Å for the Pna2$_1$ structure and $a$ = 3.408 Å, $b$ = 5.892 Å, and $c$ = 5.541 Å for the Pmc2$_1$ structure, and the difference in the total energy of the Pmc2$_1$ and Pna2$_1$ structures is 5 meV per ZnSnN$_2$ unit, which agree with previous computational results [14]. The bandgaps for Pna2$_1$- and Pmc2$_1$-ZnSnN$_2$ calculated using the PBE functional, severely underestimated, are 0.11 eV and 0.04 eV, respectively. *Mod*-HSE06 calculations predict the bandgaps for Pna2$_1$- and Pmc2$_1$-ZnSnN$_2$ to be 1.43 eV and 1.38 eV, respectively. Magnitude of the bandgap of Pna2$_1$-ZnSnN$_2$ is in good agreement with previous hybrid functional calculations. [1,16] Larger experimental bandgaps than 1.43 eV could be attributed to the Burstein-Moss effect in view of large electron carrier concentration in ZnSnN$_2$. [1]

B. Geometry and stability of "4-8"-type ZnSnN$_2$ films

Next, we consider ZnSnN$_2$ polar films terminating with the (0001)/(000$\bar{1}$) surfaces. The number of layers is calculated such that two adjacent monolayers are regarded as one layer. Figures 1(a) and 1(c) show the geometry relaxed 4-layer Pna2$_1$- and Pmc2$_1$-ZnSnN$_2$ (0001)/(000$\bar{1}$) slabs, respectively. It is visible that the surface atoms for both slabs undergo great relaxations with the top Sn atoms relaxed outward and the bottom N atoms in the Zn$_2$Sn$_1$ motif relaxed inward. Calculated electronic band structures show that both slabs have metallic characteristic (see Figures S1).

We note that below a certain thickness wurtzite polar (0001)/(000$\bar{1}$) films are not stable, while BCT films, having alternating 4- and 8-atom rings characteristic, are energetically favorable. [24,25] BCT films do not have spontaneous polarization, offering potential solutions to the internal polarization field problem for wurtzite materials. In view of the similarity between the wurtzite and Pna2$_1$ structures, we consider BCT-like structure for ZnSnN$_2$ thin films. Figures 1(b) and 1(d) present the 4-layer reconstructed slabs (about 0.94 nm) with alternating 4- and 8-atom rings transformed from the $Pna2_1$- and $Pmc2_1$-ZnSnN$_2$ polar slabs, called "4-8"-I type and "4-8"-II type slabs, respectively. 4-8 reconstruction results in a change of about 0.5% for the in-plane lattice parameters of the slabs compared to the $Pna2_1$- and $Pmc2_1$-ZnSnN$_2$ lattice parameters. We neglect the small lattice parameter changes and fix the in-plane lattice parameters of "4-8"-type slabs to the $Pna2_1$- and $Pmc2_1$-ZnSnN$_2$ ones, giving an energy accuracy within 1 meV per ZnSnN$_2$ unit for the slabs.

Our results show that for 4-layer slab the "4-8"-I (-II) phase is lower in energy by 0.452 eV (0.481 eV) per ZnSnN$_2$ unit than the $Pna2_1$ ($Pmc2_1$) phase and does not have spontaneous polarization (see Figure S2). From the top view of the slabs, the "4-8"-type structures are barely distinguishable from the $Pna2_1$ and $Pmc2_1$ structures. Table I lists the bond angles and bond lengths of 4-atom rings in the "4-8"-type slabs. The bond angles N-Sn-N and N-Zn-N (denoted by $\theta_1$ and $\theta_2$, respectively) are $\theta_1 = 88.06°$ and $\theta_2 = 88.84°$ for the "4-8"-I slab in Fig. 1(b) and $\theta_1 = 87.80°$ and $\theta_2 = 90.68°$ for the "4-8"-II slab in Fig. 1(d), indicating distorted tetrahedral coordination. The bond lengths Sn-N (denoted by $d_1$ and $d_2$) and Zn-N (denoted by $d_3$ and $d_4$) for the "4-8"-I slab are $d_1 = 2.186$ Å, $d_2 = 2.134$ Å, $d_3 = 2.119$ Å, and $d_4 = 2.093$ Å, which are close to those for the "4-8"-II slab with the same trend that the length of out-of-plane bond is slightly larger than that of in-plane bond, i.e., $d_1 > d_2$ and $d_3 > d_4$.

To examine the energy stability of "4-8"-type slabs as the total number of layers increases, we calculate the energies per ZnSnN$_2$ unit of the slabs for 4-8, $Pna2_1$, and $Pmc2_1$ phases, as shown in Fig. 2. The energies for $Pna2_1$ and $Pmc2_1$ slabs decrease with increasing the number of layers, while the energies for "4-8"-type slabs are subject to considerable fluctuation with even-layer slabs being lower in energy than adjacent odd-layer ones. We note that Sn's 5s and 5p orbitals have four electrons and in even-layer slabs all Sn atoms are in four-fold coordination, while in odd-layer slabs top Sn atoms are in three-fold coordination. Different bonding properties for even- and odd-layer "4-8"-type slabs render odd-layer slabs higher in energy than adjacent even-layer slabs. We compare the energies of even-layer "4-8"-type slabs with those of polar slabs, finding that even-layer "4-8"-I (-II) slabs with thickness below 14 (16) layers have lower energies than $Pna2_1$ ($Pmc2_1$) polar slabs.

Transition pathways from the $Pna2_1$/$Pmc2_1$ to "4-8"-type structures are calculated using the climbing image nudged elastic band method [33], as shown in Fig. 3. From the energy profiles, the energy barrier for the $Pna2_1$ ($Pmc2_1$) to "4-8"-I (-II) transition for 4-layer slab is estimated to be 0.104 eV (0.094) per unit cell. The small energy barriers indicate that the structural transition for 4-layer slabs can spontaneously occur at finite temperatures. We perform additional molecular dynamics simulations at 300 K in the canonical ensemble with 4-layer $Pna2_1$- and $Pmc2_1$-$ZnSnN_2$ (0001)/(000$\bar{1}$) slabs as the initial structures (see Figures S3 and S4). Our results show that for 4-layer slabs the $Pna2_1$ and $Pmc2_1$ structures are spontaneously transformed to the "4-8"-type structure. For 6-layer slabs, however, the energy barrier for the $Pna2_1$ ($Pmc2_1$) to "4-8"-I (-II) transition is calculated to be 1.325 eV (0.664 eV) per unit cell, which is one order of magnitude larger than that for the 4-layer slabs.

C. Electronic structure of "4-8"-type ZnSnN$_2$ slabs

Figures 4(a)-(d) present the calculated electronic band structure and orbital-resolved density of states (DOS) for 4-layer "4-8"-type slabs. All slabs show a semiconducting nature, dramatically different from the metallic property for polar slabs. For 4-layer "4-8"-I slab, the bandgap is quasi-direct with a value of 0.57 eV at the PBE level of theory between the conduction band minimum (CBM) at the Γ point and the valence band maximum (VBM) along the Γ to X direction, near the Γ point. The difference between the bandgap at Γ and the quasi-direct bandgap is tiny (< 1 meV). For 4-layer "4-8"-II slab, the bandgap is direct with the CBM and VBM located at Γ and is 0.41 eV at the PBE level of theory. The orbital-resolved DOS indicates that for both slabs the VBM is dominated by N 2p states and the upper valence band consists of hybridization of N 2p and Zn 3d states, while the CBM is mainly contributed by N 2s and Sn 5s states.

The bandgaps for "4-8"-type slabs decrease as the number of layers increases, as shown in Tables 2 and 3. The *mod*-HSE06 calculations exhibit similar trend compared with the PBE results but give more accurate bandgaps for the slabs studied, ranging from 1.4 eV to 1.8 eV. The electron effective mass of the conduction band at Γ in the *x* direction is close to that in the *y* direction due to the prominent *s*-orbital property of the CBM. Small values of electron effective masses indicate probably high electron mobility of these slabs.

"4-8"-type ZnSnN$_2$ thin films have potential applications in visible light responsive optoelectronic devices due to the suitable bandgaps. Given the colossal computational cost for the hybrid functional optical calculation based on dense **k**-point grid, we only calculate the optical spectrum for 4-layer "4-8"-II slab using the hybrid functional and 20×12×1 Γ-centered **k**-point grid, as shown in Fig. 5. It is observed that the optical absorption is quite strong over the visible light energy range.

D. Disorder effect

The "4-8"-I and "4-8"-II phases have close energies with the latter slightly lower in energy by 21, 24, 26, and 26 meV per $ZnSnN_2$ unit for 4-, 6-, 8-, and 10-layer slabs, respectively, implying the likely occurrence of the mixing of two phases under the experimental conditions. Quayle *et al*. showed that the $Pna2_1$ and $Pmc2_1$ structures of $ZnSnN_2$ can be viewed as stacking of ±1 pseudospin layers. [14] As shown in APPENDIX, the $Pna2_1$ structure arises from the +1, -1 stacking along the $Pna2_1$ **b** axis and the $Pmc2_1$ structure arises from the -1, -1 or +1, +1 stacking. It is worth noting that the formation of octet-rule-violating disordered structure needs high energetic cost, while the octet-rule-preserving disordered structure can be readily formed. [14] Thus, using the (1×3) supercell of 4-layer "4-8"-type ordered slab, we build an octet-rule-preserving disordered structure by swapping neighboring pseudospin layers, as shown in Fig. 6, to study the disorder effect on the properties of "4-8"-type slab.

The energy per $ZnSnN_2$ unit of disordered phase is lower by 6 meV than that of "4-8"-I phase and higher by 15 meV than that of "4-8"-II phase. Figure 7(a) shows the calculated electronic band structure of 4-layer "4-8"-type disordered slab, the bandgap of which is direct and is 0.53 eV at the PBE level of theory that lies between the bandgaps of "4-8"-I and "4-8"-II phases. The calculated electron effective masses are 0.13 $m_e$ (0.12 $m_e$) in the $x$ ($y$) direction at the PBE level of theory, which are the same as those of "4-8"-I ordered slab. Figure 7(b) shows the band-decomposed charge densities for the VBM and CBM states. The VBM state is largely distributed in the inner layers of slab, while the CBM state is almost uniformly distributed in all layers of slab. These results indicate that octet-rule-preserving disorder has insignificant effects on the electronic properties of "4-8"-type slabs.

## IV. Conclusions

In this work, "4-8"-type ZnSnN$_2$ thin films are predicted using first-principles calculations. Below a certain thickness, "4-8"-type thin films have lower total energy than polar $(0001)/(000\bar{1})$ films. Furthermore, for "4-8"-type films, even-layer ones are lower in energy than adjacent odd-layer ones. The energy barrier for the Pna2$_1$/Pmc2$_1$ → 4-8 transition depends on the thickness of films. For 4-layer thin film, the barrier is about 0.1 eV per unit cell and thus the transition can spontaneously occur at finite temperatures. All "4-8"-type thin films studied are non-polar and semiconducting with the bandgaps ranging from 1.4 eV to 1.8 eV depending on the thickness of films. These films show light electron effective masses that are hardly affected by the octet-rule-preserving disorder. Our work paves the way for future experimental synthesis and optoelectronic applications of "4-8"-type ZnSnN$_2$.

**Supporting Material**

Electronic band structure for 4-layer Pna2$_1$- and Pmc2$_1$-ZnSnN$_2$ $(0001)/(000\bar{1})$ slabs; planar-averaged electrostatic potential; molecular dynamics simulations at 300 K; "4-8"-type disordered slab based on (1×2) supercell.

**Conflicts of interest**

There are no conflicts of interest to declare.

**Acknowledgments**

We acknowledge the financial support from the National Natural Science Foundation of China (Grant No. 11604254) and the Natural Science Foundation of Shaanxi Province (Grant No. 2019JQ-240). We also acknowledge the HPCC Platform of Xi'an Jiaotong University for providing the computing facilities.

Table 1 Bond angles ($\theta_i$) and bond lengths ($d_i$) (in Å) for 4-layer "4-8"-I and "4-8"-II slabs as shown in Figs. 1(b) and 1(d).

| Configurations | $\theta_1$ | $\theta_2$ | $d_1$ | $d_2$ | $d_3$ | $d_4$ |
|---|---|---|---|---|---|---|
| "4-8"-I | 88.06° | 88.84° | 2.186 | 2.134 | 2.119 | 2.093 |
| "4-8"-II | 87.80° | 90.68° | 2.174 | 2.148 | 2.118 | 2.096 |

Table 2 Bandgap $E_g$ (in units of eV) and electron effective mass (in units of free electron mass $m_e$) calculated using the PBE functional for even-layer "4-8"-I slabs and 4-layer "4-8"-type disordered slab. The values in parentheses correspond to the *mod*-HSE06 results.

| Number of layers | $E_g$ | $m_x$ | $m_y$ |
| --- | --- | --- | --- |
| 4 | 0.57 (1.84) | 0.13 (0.18) | 0.12 (0.17) |
| 4 (disordered) | 0.53 | 0.13 | 0.12 |
| 6 | 0.44 (1.70) | 0.11 (0.17) | 0.10 (0.16) |
| 8 | 0.36 (1.63) | 0.10 (0.16) | 0.09 (0.15) |
| 10 | 0.31 (1.58) | 0.09 (0.15) | 0.09 (0.15) |

Table 3 Bandgap $E_g$ (in units of eV) and electron effective mass (in units of free electron mass $m_e$) calculated using the PBE functional for even-layer "4-8"-II slabs. The values in parentheses correspond to the *mod*-HSE06 results.

| Number of layers | $E_g$ | $m_x$ | $m_y$ |
|---|---|---|---|
| 4 | 0.41 (1.69) | 0.11 (0.16) | 0.14 (0.18) |
| 6 | 0.28 (1.56) | 0.09 (0.15) | 0.11 (0.16) |
| 8 | 0.21 (1.48) | 0.08 (0.14) | 0.10 (0.16) |
| 10 | 0.17 (1.44) | 0.08 (0.14) | 0.09 (0.15) |

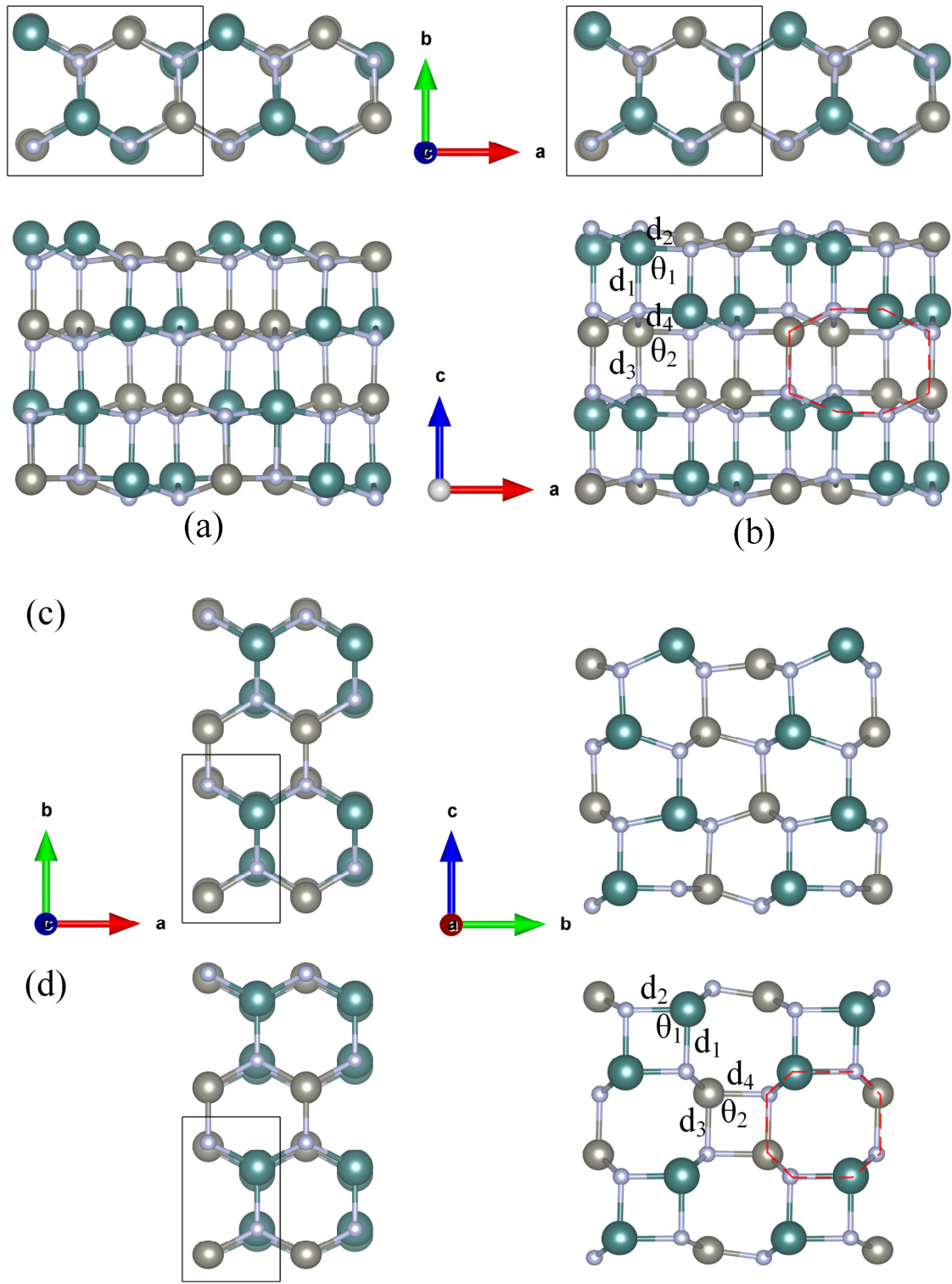

Fig. 1 Top and side views of 4-layer ZnSnN$_2$ slabs in the Pna2$_1$ (a) and "4-8"-I (b) structures and in the Pmc2$_1$ (c) and "4-8"-II (d) structures. Small balls are N atoms, and large gray and teal balls are Zn and Sn atoms, respectively. 8-atom rings are indicated by red dashed lines. Primitive cells are denoted by the black frames. Bond angles are denoted by $\theta_i$ (i = 1 – 2), and bond lengths, $d_i$ (i = 1 – 4).

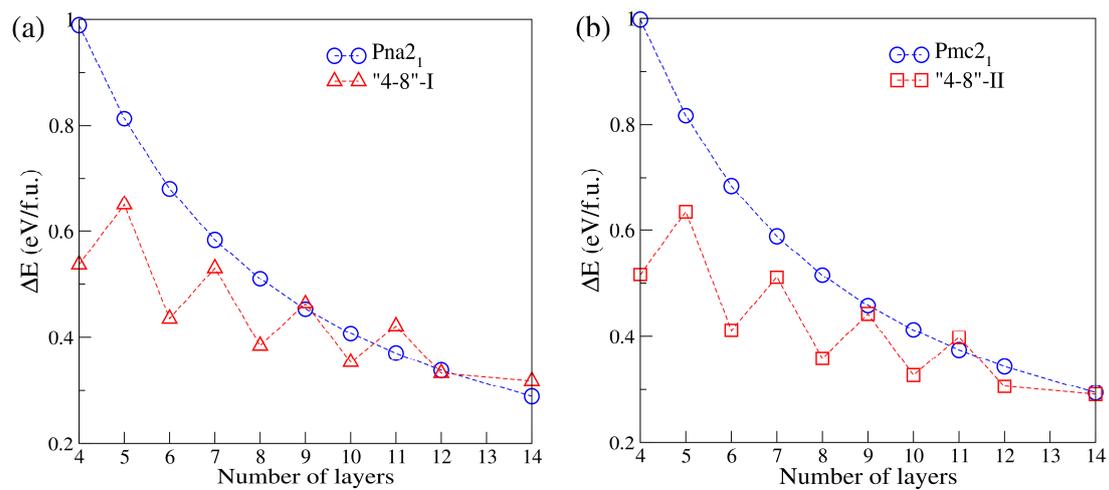

Fig. 2 Energies per formula unit (f.u.) relative to bulk $Pna2_1$-$ZnSnN_2$ as a function of the number of layers for the $Pna2_1$ and "4-8"-I $ZnSnN_2$ films (a) and for the $Pmc2_1$ and "4-8"-II $ZnSnN_2$ films (b).

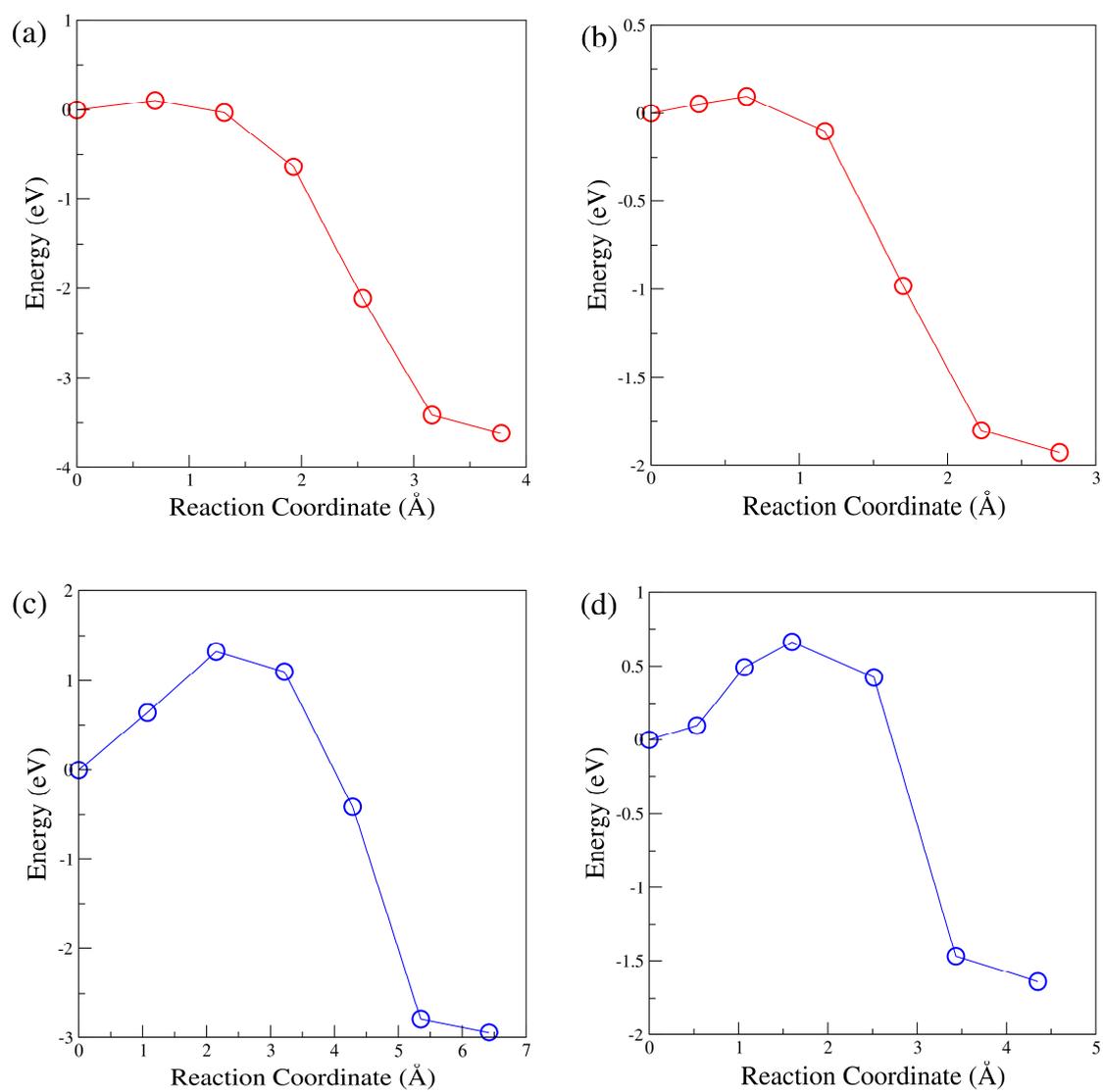

Fig. 3 Energy pathways for 4-layer slabs from the Pna2$_1$ to "4-8"-I structures (a) and from the Pmc2$_1$ to "4-8"-II structures (b) and for 6-layer slabs from the Pna2$_1$ to "4-8"-I structures (c) and from the Pmc2$_1$ to "4-8"-II structures (d). The start structures are the optimized Pna2$_1$/Pmc2$_1$ structures and the end structures are the optimized "4-8"-type structures.

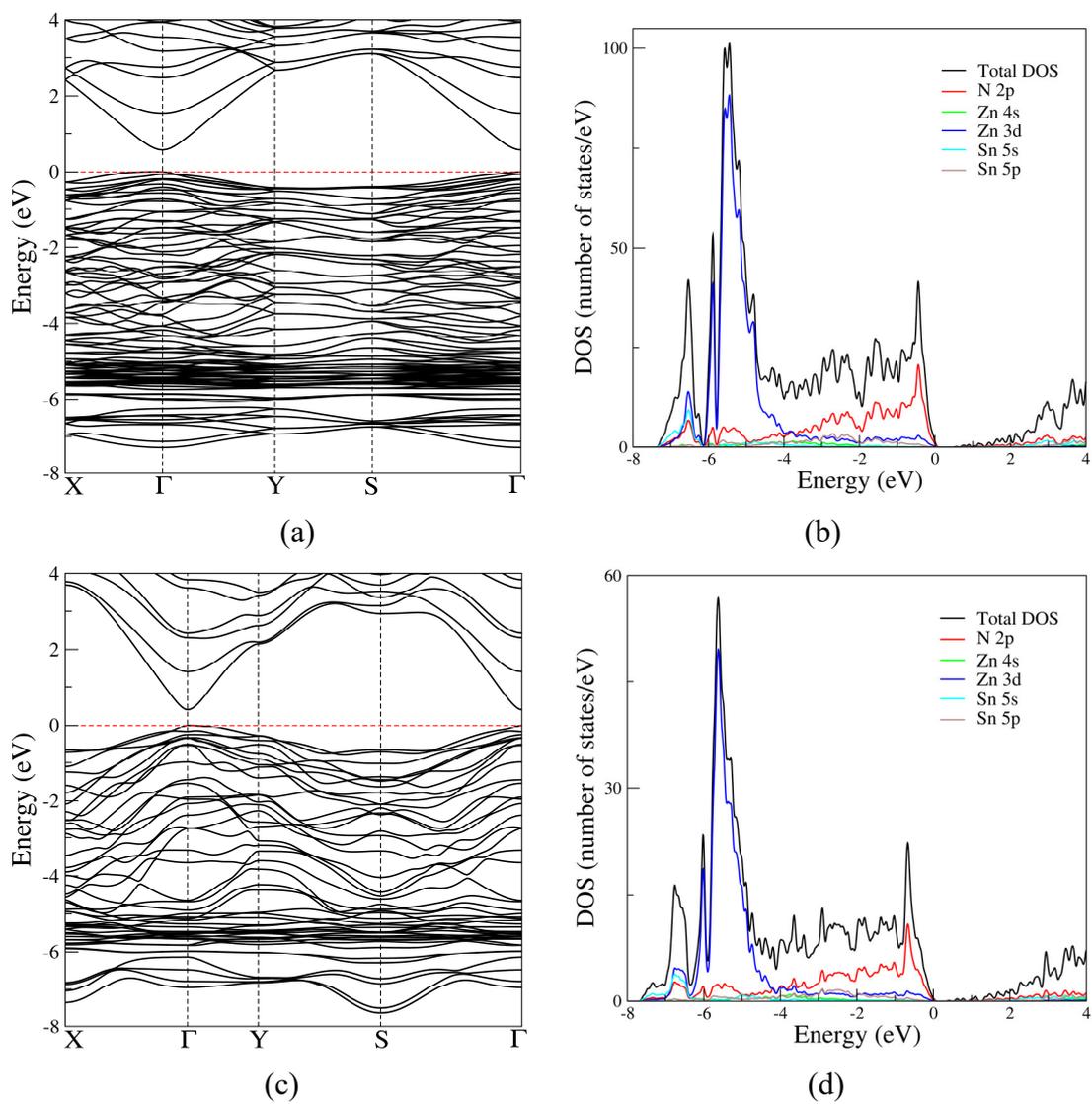

Fig. 4 Electronic band structure and orbital-resolved density of states (DOS) calculated using the PBE functional for 4-layer "4-8"-I [(a) and (b)] and "4-8"-II [(c) and (d)] slabs.

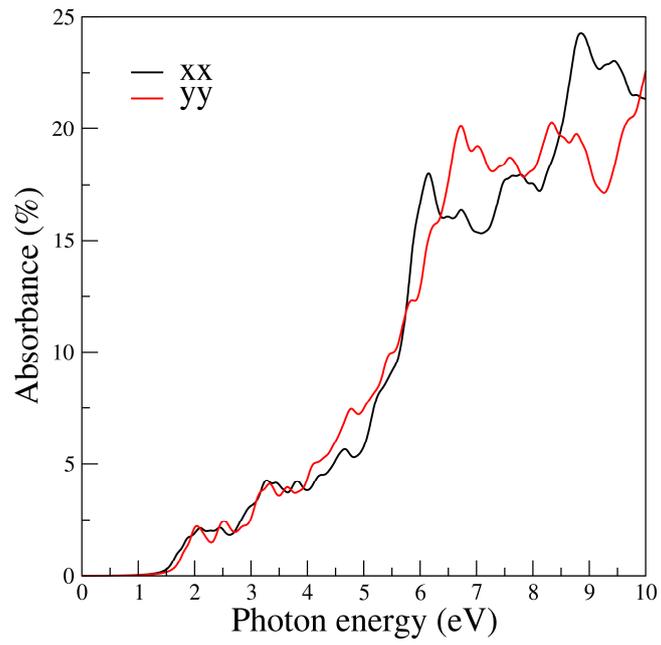

Fig. 5 xx and yy components of the absorption spectrum for 4-layer "4-8"-II slab calculated using the hybrid functional and 20×12×1 Γ-centered **k**-point grid.

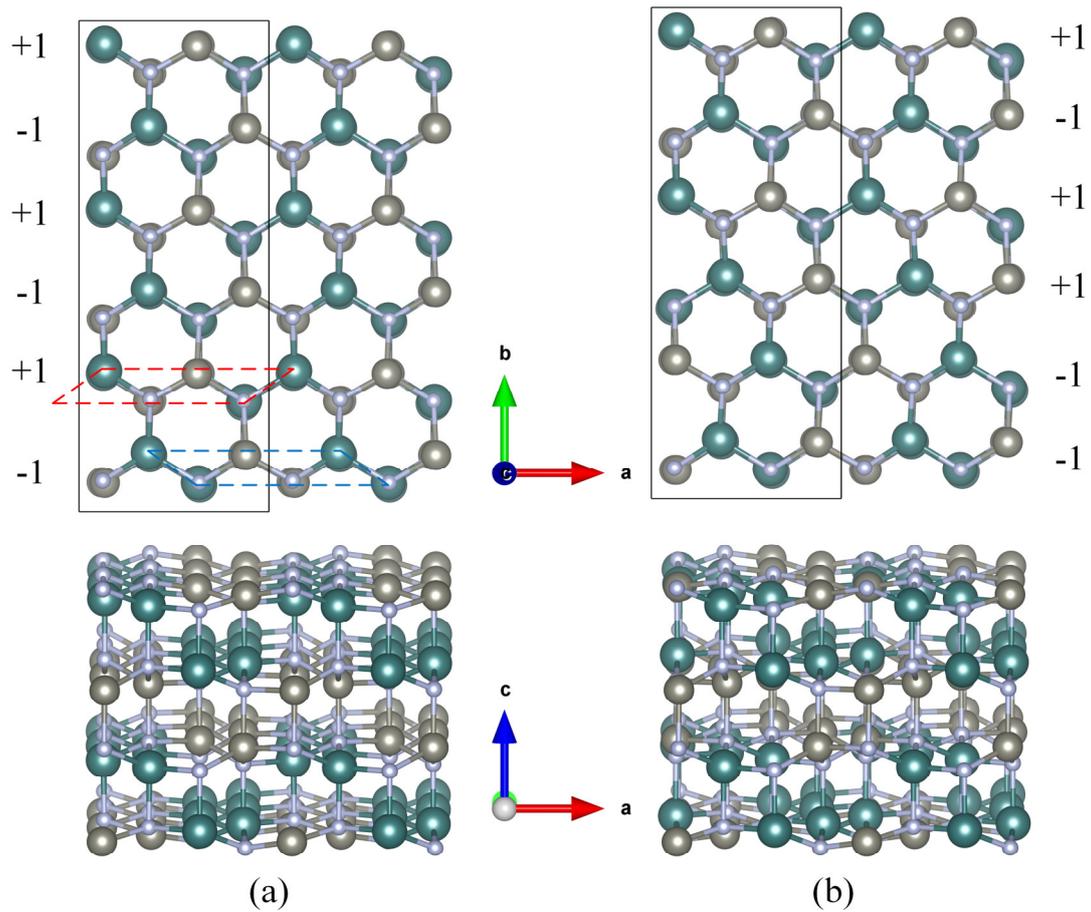

Fig. 6 Top and side views of the (1×3) supercells (denoted by the black frame) of 4-layer "4-8"-type ordered (a) and disordered (b) slabs.

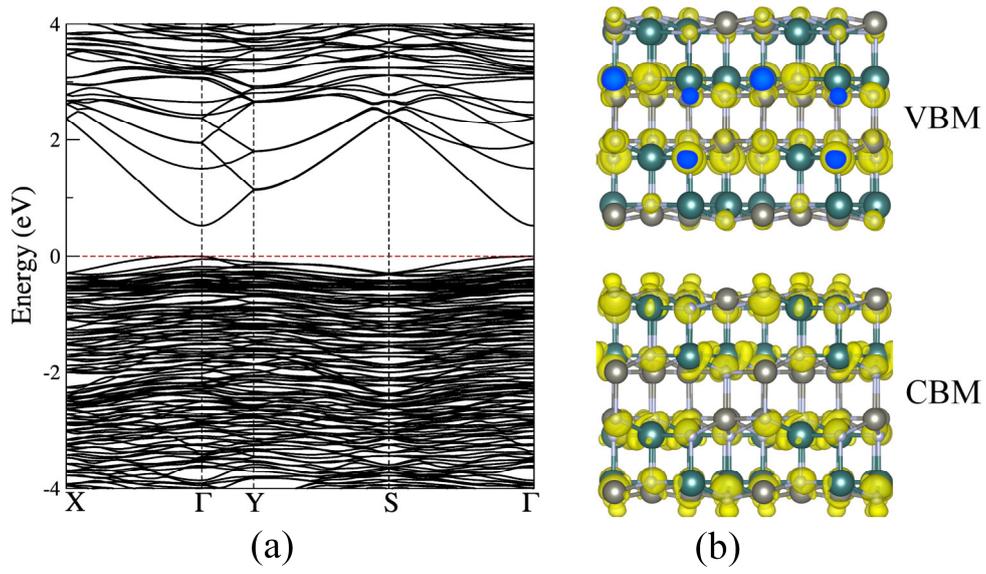

Fig. 7 Electronic band structure (a) and band-decomposed charge densities for the VBM and CBM states (b) for 4-layer "4-8"-type disordered slab.

APPENDIX.

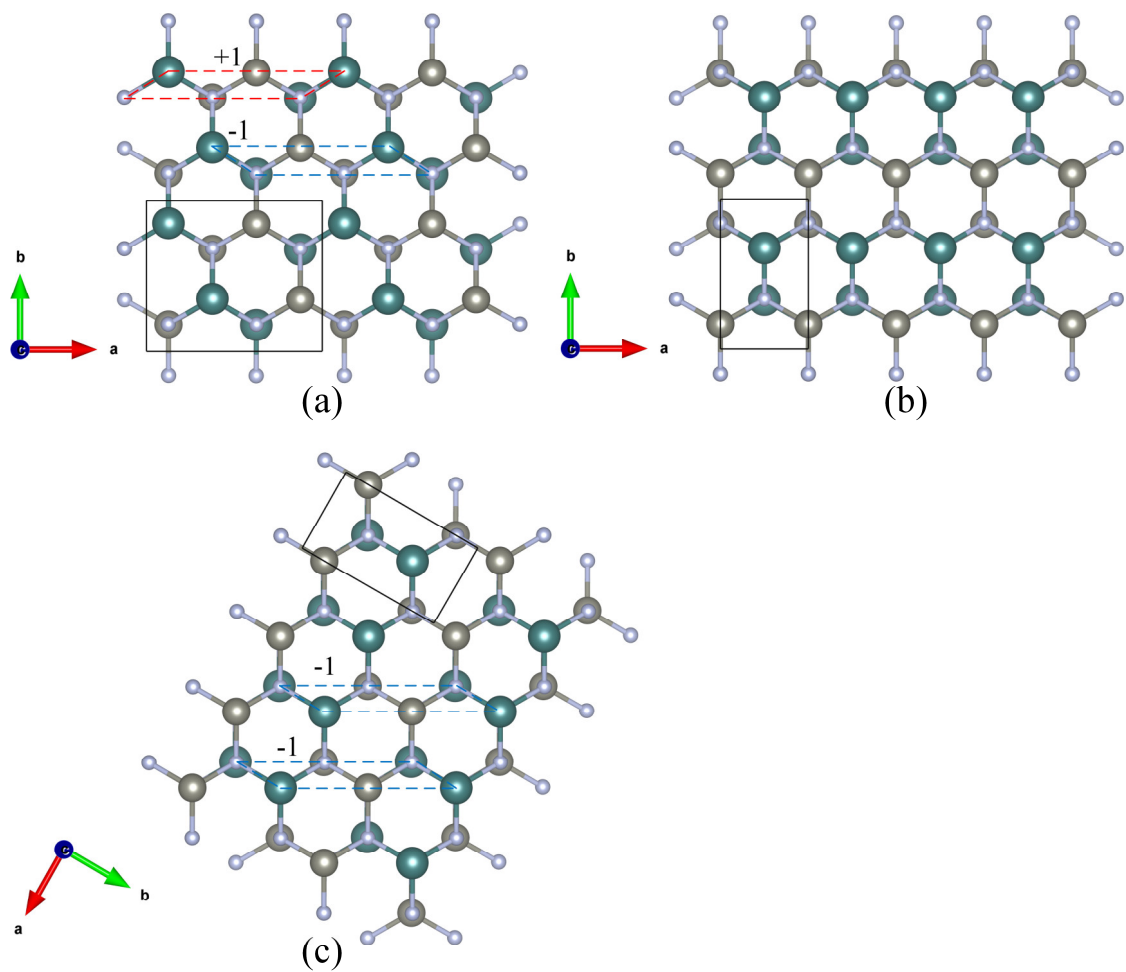

Fig. 8 Crystal structures of bulk ZnSnN$_2$ with the Pna2$_1$ (a) and Pmc2$_1$ (b) space groups. (c) The Pmc2$_1$ crystal structure rotated in clockwise direction by 120° with respect to panel (b). Primitive cells are denoted by the black frames. In panel (a), +1 and -1 pseudospin layers are outlined in dashed red and blue, respectively.

**Supporting material:**

**Stability and electronic properties of "4-8"-type ZnSnN$_2$ thin films free of spontaneous polarization for optoelectronic devices**

D. Q. Fang[1*]

[1]MOE Key Laboratory for Nonequilibrium Synthesis and Modulation of Condensed Matter, School of Physics, Xi'an Jiaotong University, Xi'an 710049, China

*fangdqphy@xjtu.edu.cn

Figures S1-S5.

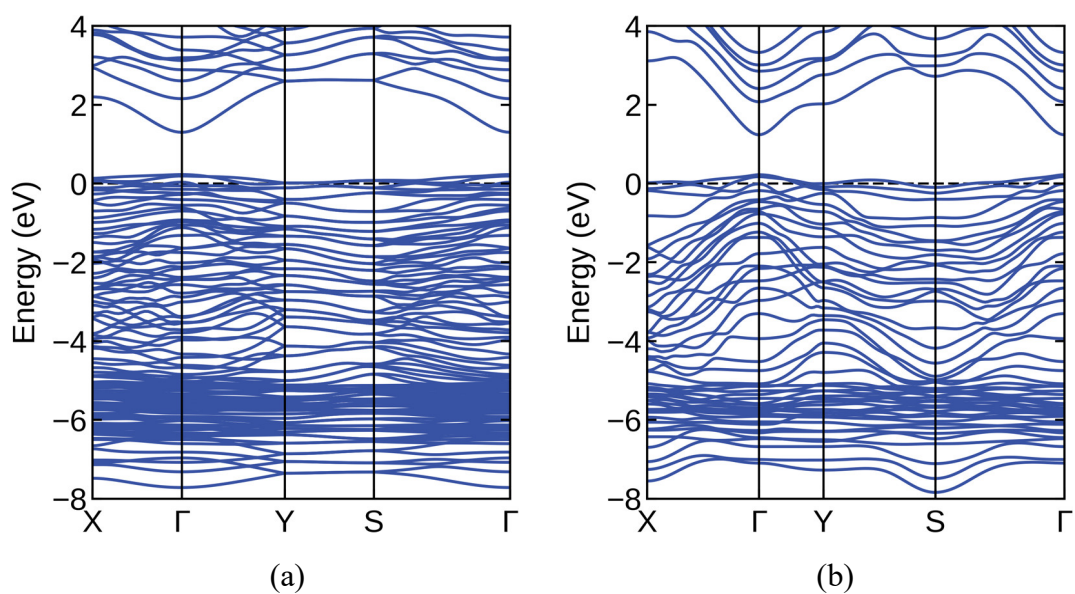

Figure S1. Electronic band structure for 4-layer Pna2$_1$- (a) and Pmc2$_1$-ZnSnN$_2$ (b) (0001)/(000$\bar{1}$) slabs calculated using the PBE functional. The Fermi energy is set to zero.

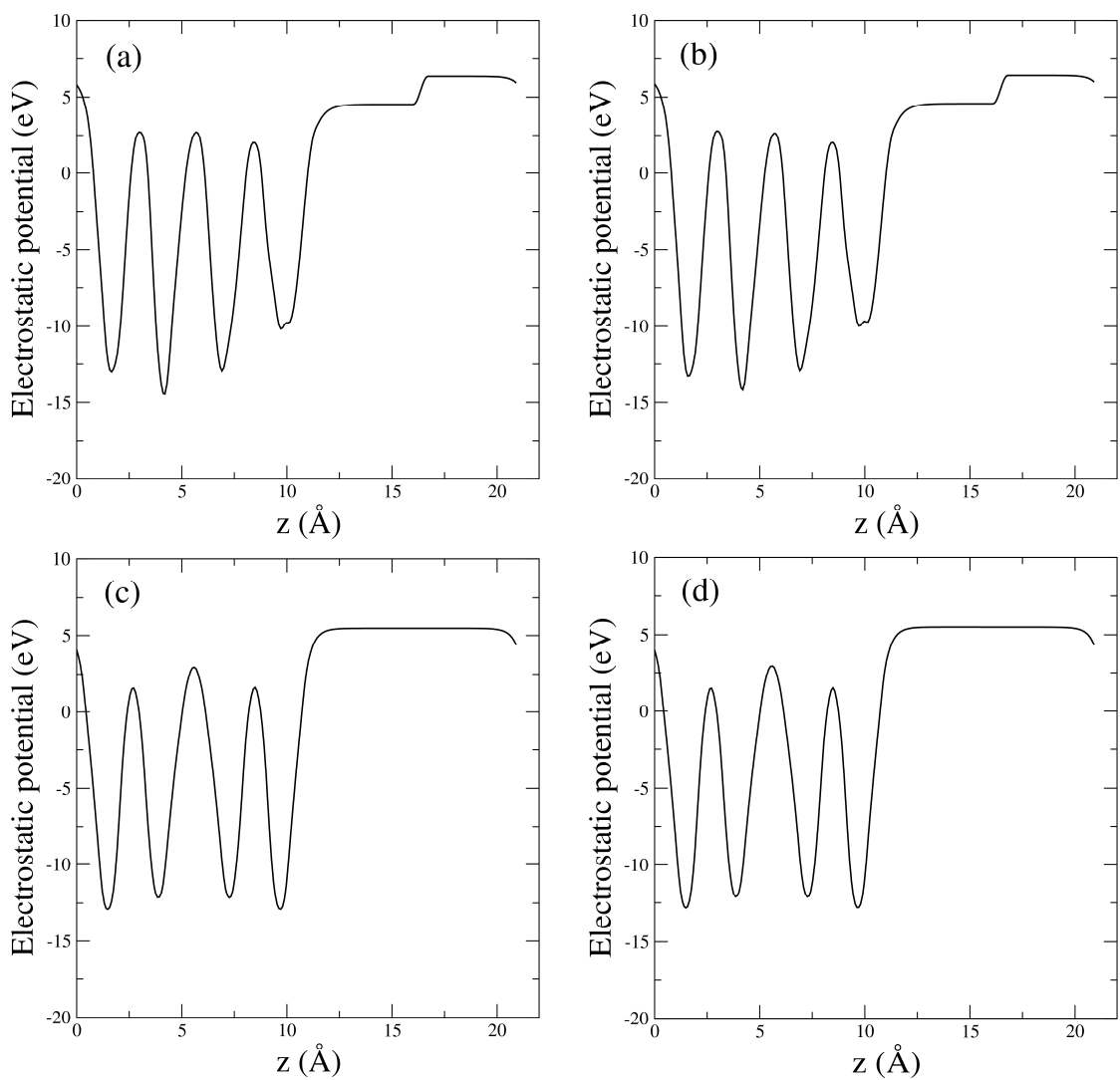

Figure S2. Planar-averaged electrostatic potential for 4-layer Pna2$_1$- (a) and Pmc2$_1$-ZnSnN$_2$ (b) (0001)/(000$\bar{1}$) slabs and 4-layer "4-8"-I (c) and "4-8"-II (d) slabs.

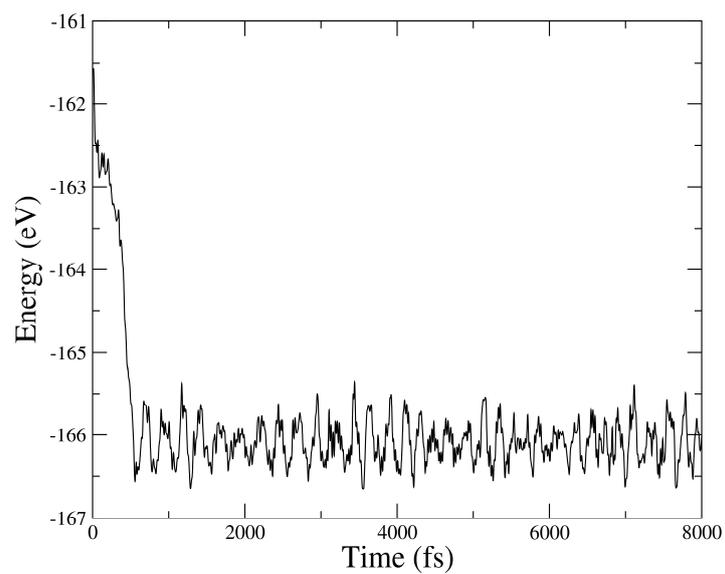

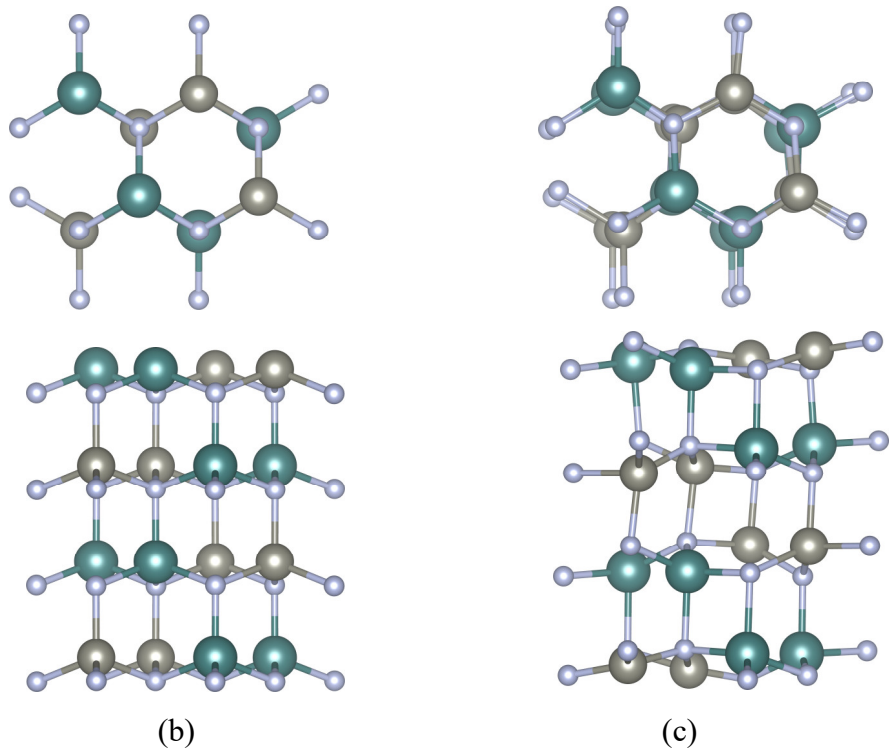

Figure S3. Energy variation of 4-layer Pna2$_1$-ZnSnN$_2$ (0001)/(000$\bar{1}$) slab with time obtained from the molecular dynamics simulation at 300 K in a canonical ensemble (a). Top and side views of the initial structure are shown in (b), and the structure at the end of 8 ps in (c).

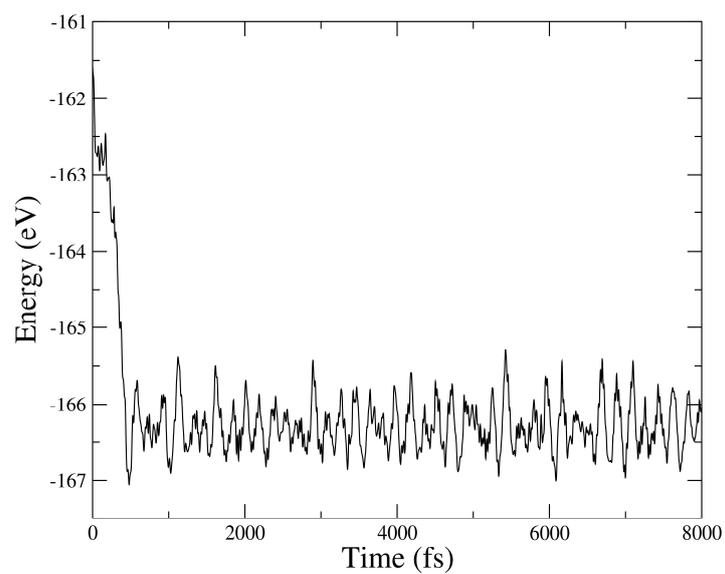

(a)

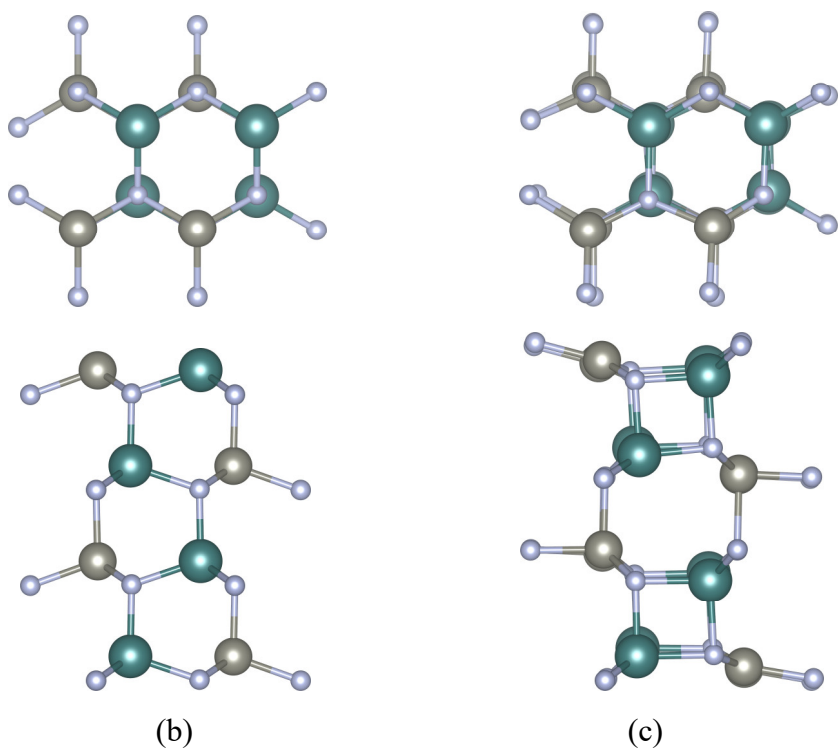

(b)          (c)

Figure S4. Energy variation of 4-layer Pmc2$_1$-ZnSnN$_2$ (0001)/(000$\bar{1}$) slab with time obtained from the molecular dynamics simulation at 300 K in a canonical ensemble. Top and side views of the initial structure are shown in (b), and the structure at the end of 8 ps in (c).

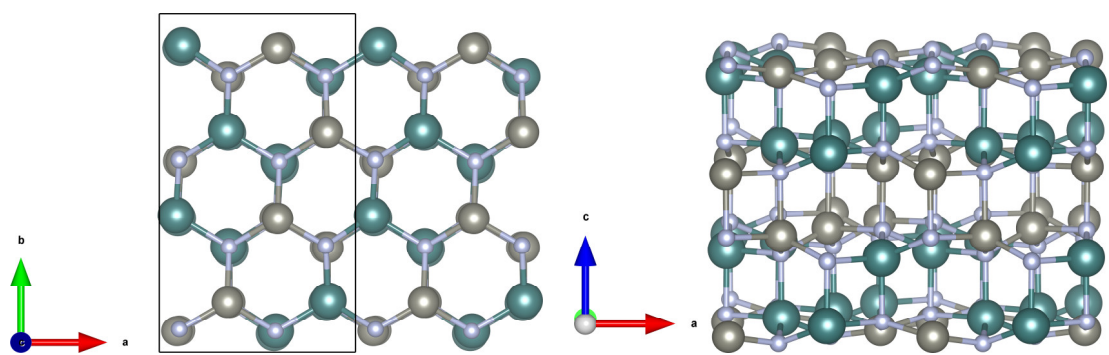

(a)

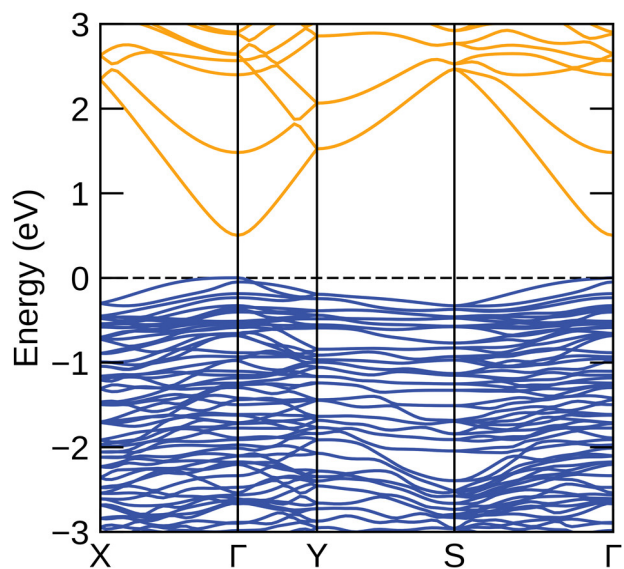

(b)

Figure S5. Top and side views of the "4-8"-type disordered slab based on (1×2) supercell that is denoted by the black frame (a) and its electronic band structure. This disordered slab has a direct bandgap of 0.50 eV with an electron effective mass of 0.13 $m_e$ (0.12 $m_e$) in the $x$ ($y$) direction at the PBE level of theory.